 \documentclass[final,3p,times]{elsarticle}
\usepackage{graphicx}
\usepackage{amssymb}
\journal{XXX}

\begin{document}

\begin{frontmatter}


\title{{A Bayesian approach for extracting free energy profiles from cryo-electron microscopy experiments using a path collective variable}}


\author[1,2,+]{Julian Giraldo-Barreto}
\author[1,+]{Sebastian Ortiz}
\author[3]{Erik H. Thiede}
\author[4]{Karen Palacio-Rodriguez}
\author[3]{Bob Carpenter}
\author[3]{Alex H. Barnett}
\author[1,5,*]{Pilar Cossio}

\address[1]{ \normalsize {\it Biophysics of Tropical Diseases Max Planck Tandem Group, University of Antioquia, Medellín, Colombia.}}
\address[2]{\normalsize \textit{Magnetism and Simulation Group, University of Antioquia, Medellín, Colombia.} }
\address[3]{ \normalsize {\it Center for Computational Mathematics, Flatiron Institute, New York City, United States of America.}}
\address[4]{\normalsize {\it Sorbonne Université, Institut de Minéralogie, de Physique des Matériaux et de Cosmochimie, Paris, France.}}
\address[5]{\normalsize \textit{Department of Theoretical Biophysics, Max Planck Institute of Biophysics, 60438 Frankfurt am Main, Germany.}}

\address[*]{{\small\textit{email:} pilar.cossio@biophys.mpg.de; grupotandem.biotd@udea.edu.co}}
\address[+]{Equal contribution.}
%\affil[$\ddagger$]{{\small Equal contribution.}}

\begin{abstract}
Cryo-electron microscopy (cryo-EM) extracts single-particle density projections of individual biomolecules.
Although cryo-EM is widely used for 3D reconstruction, due to its single-particle nature, it has the potential to provide information about the biomolecule's conformational variability and underlying free energy landscape. 
However, treating cryo-EM as a single-molecule technique is challenging because of the low signal-to-noise ratio (SNR) in the individual particles.
In this work, we developed the cryo-BIFE method, cryo-EM Bayesian Inference of Free Energy profiles, that uses a path collective variable to extract free energy profiles and their uncertainties from cryo-EM images. 
We tested the framework over several synthetic systems, where we controlled the imaging parameters and conditions. We found that for realistic cryo-EM environments and relevant biomolecular systems, it is possible to recover the underlying free energy, with the pose accuracy and SNR as crucial determinants.  Then, we used the method to study the conformational transitions of a calcium-activated channel with real cryo-EM particles. Interestingly, we recover the most probable conformation (used to generate a high resolution reconstruction of the calcium-bound state), and we find two additional meta-stable states, one which corresponds to the calcium-unbound conformation. As expected for turnover transitions within the same sample, the activation barriers are of the order of a couple $k_BT$.   
Extracting free energy profiles from cryo-EM will enable a more complete characterization of the thermodynamic ensemble of biomolecules. 
\end{abstract}

\begin{keyword}

 cryo-EM \sep free energy \sep collective variable \sep Bayesian \sep path \sep individual particles

\end{keyword}

\end{frontmatter}

%\linenumbers

\section*{Introduction}
In cryo-electron microscopy (cryo-EM) experiments, 2D projection images are taken of a biomolecular sample immersed in vitrified ice and irradiating it with a low electron dose. Due to advances in the electron detection cameras~\citep{McMullan2016} and improvements in the reconstruction algorithms~\citep{Cossio:COSB:2018}, cryo-EM now enables resolving density maps at near-atomic resolution ~\citep{Cheng2015} with highest resolution close to 1.22 \AA \citep{Stark2020,Nakane2020}. Therefore, cryo-EM now plays a principal role in structural biology for understanding biological systems of a wide range of sizes (from a few KDa to a hundreds of MDa)~\citep{Murata2018}.

A main difference---and advantage---of cryo-EM with respect to X-Ray crystallography is that the vitreous ice solution can contain molecules in diverse configurational states. The ultra-fast vitrification process \citep{Dubochet1988} traps the biomolecules in the configurations they take at the temperature before flash-cooling, and the conformational ensemble follows Boltzmann's distribution. 
The absence of a single rigid crystalline structure is of great advantage to study the thermodynamic ensemble of the biomolecule~\citep{Lederman2020,Murata2018,Frank2016}. With cryo-EM it is not only possible to extract heterogeneous 3D conformations of the system but, in principle, one can characterize relevant biophysical properties, such as the free energy landscape, activation barriers, transition states and transition paths between the conformations. This can provide essential clues about the biomolecule's function \citep{Frank2016}.

Several methods have been developed to extract 3D density maps of heterogeneous biomolecules from cryo-EM. The methods can be divided in two types: discrete-state or continuous-state methods. Discrete methods start from a discrete set of reference maps and classify the cryo-EM images according to the map they are more alike to. The classified subsets are optimized iteratively during the refinement \citep{Scheres2012,Grigorieff2016,Punjani2017}. However, these approaches may be biased towards the initial maps used as templates, and the number of discrete classes has to be predetermined \citep{Jonic2017}. To overcome some of these limitations statistical analysis methods that use principal component analysis (PCA)~\citep{Penczek2011,Tagare2015}, normal mode analysis~\citep{Jin2014} or the covariance matrix~\citep{Liao2015,Katsevich2015,Anden2018} have been developed. Along these lines, combining statistical analysis with optimization algorithms can result in more efficient methods to reconstruct 3D density maps \citep{Punjani2020.04.08.032466,Lederman2020,Zhong2019reconstructing}. However, it is not trivial to determine if the system's conformational changes are best modeled by a discrete or continuous set of states \citep{Jonic2017}.

The first studies that extracted free energies directly from cryo-EM experiments used particle-classification tools. These focused on the prototypical Brownian machine: the ribosome, which uses thermal energy for functioning. Fischer and co-workers \citep{Fischer2010} characterized the free energy landscape of the slow back-translocation process using the number of classified particles to each sub-state ($n_i$;~   \textit{i.e.}, the occupancy or population of state $i$). The free energy difference with respect to a reference state ($\Delta G$) with population $n_o$ is extracted using the Boltzmann factor, $n_i/n_o = \exp(-\beta\Delta G)$, where $\beta=1/(k_BT)$ and $k_B$ is Boltzmann's constant and $T$ is the temperature. Interestingly, the authors found a relatively flat energy landscape projected along the 30S head versus body rotation at ambient temperature. A similar analysis was also applied to study a pretranslocational mRNA-tRNA sample as a function of the inter-subunit rotation angle \citep{Agirrezabala2012}.
However, these studies have the limitation of using a small number of 3D classes or relying on time information from the back-translocation process \citep{Fischer2010}.

An alternative methodology, which also studied the ribosome as a Brownian machine, was developed by Dashti and co-workers  ~\citep{Dashti2014} to extract free energies using the raw cryo-EM particles with diffusion maps. The method selects the images belonging to the same projection direction, then projects the multidimensional free energy landscape onto a low-dimensional manifold. This method has the advantage that it uses just the raw images without requiring prior 3D classes. Seitz and Frank ~\citep{Seitz2020} use this method together with the POLARIS approach for finding the least action path from 2D energy surfaces. Dashti and co-workers~\citep{Dashti2020} extracted the free energy surfaces of the ryanodine receptor type 1 (RyR1) associated with the binding-unbinding states (with the ATP, caffeine and $CA^{2}$~ligands) using a master equation approach to find the probability of a transition between the two free energy landscapes. Recently, deep learning methods have provided similar strategies to extract free energy surfaces \citep{Wu2020.12.22.423932,chen2021deep}. We note that replicating these methods might be cumbersome, and the bank of images required is very large. Moreover, the low-dimensional space upon which the particles are projected can be difficult to interpret. 

For these reasons, some recent studies have returned to particle-classification schemes for extracting free energies but they have increased the number of 3D conformations used in the classification. Haselbach and co-workers \citep{Haselbach2018} studied the dynamics of the Human Spliceosomal B$^{act}$ Complex by performing PCA on the reconstructed 3D volumes. The population of each sub-state along the first two PCA eigenvectors was used to extract the free energy landscape using the Boltzmann factor. A different study assessed the motion of unliganded glutamate dehydrogenase \citep{Oide2020}, using a hybrid approach that combined PCA over a molecular dynamics (MD) trajectory (to define the low-dimensional space) with the populations of four cryo-EM maps. The weights of the MD conformations and the relative occupancy of the particles were combined to produce a hybrid free energy landscape. These methods have the advantage of mapping the free energy onto an easy-to-interpret low-dimensional space. However, PCA assumes that the motions can be modeled in a linear regime, which might not be the case for large conformational changes. Moreover, for highly flexible molecules, generating the 3D maps might be challenging.

In contrast to particle-classification schemes, free energy profiling by means of reaction coordinates or collective variables (CV) has been widely used for understanding biomolecular processes, such as those extracted from MD simulations.
CVs reduce the dimensionality of the system by projecting the molecular coordinates onto a low-dimensional continuous variable (note that PCA is a particular method for constructing CVs). Adequate CVs provide information about how a reaction takes place and how relevant conformational changes occur, \textit{e.g.}, in the analysis of protein folding. A good CV should be able to discriminate the relevant metastable states of the system, and the projected free energy profile should have the key points from the underlying multidimensional free energy (such as the barrier heights and transition states). Free energies are commonly extracted by evaluating the CV for each conformation, taking a histogram of the values and relating the population of each bin to the free energy using the Boltzmann factor. However, approaches based on Bayesian approaches also exist~\citep{Stecher2014}. CVs have also been used with enhanced sampling techniques, such as umbrella sampling~\citep{Torrie1977} or metadynamics~\citep{Laio2002} that bias the simulation along the CVs in order to more efficiently explore the conformational space for extracting the free energy landscape. CVs provide a simple and continuous low-dimensional projection of the free energy landscape of a complex multidimensional systems. Along these lines, several methods \citep{Bonomi2018,Vant2020}  have been proposed to extract free energies from MD simulations with CVs that use 3D  cryo-EM maps instead of directly using the individual particles.

Inspired by the CVs from the MD community~\citep{Branduardi2007}, and motivated by extracting free energies directly from cryo-EM particle images~\citep{Dashti2014}, we propose the cryo-BIFE method (cryo-EM Bayesian Inference of Free Energy profiles) that uses a simple path CV together with a Bayesian formalism for extracting free energy profiles and their uncertainties from cryo-EM particles. We apply the method over several datasets that represent a diverse set of biomolecular systems using controlled parameters and known underlying free energy profiles. We show that under several realistic cryo-EM conditions it is possible to recover the free energy profile using our methodology. We then apply it with real cryo-EM data to study the transition between the calcium bound/unbound states of a membrane channel. We expect that free energy profiles from cryo-EM particles will bring new information about the metastable states, barriers, and transition states, to obtain a more complete thermodynamic characterization of the biomolecular system.

\begin{figure}[!t]
\centering
\includegraphics[width=12cm]{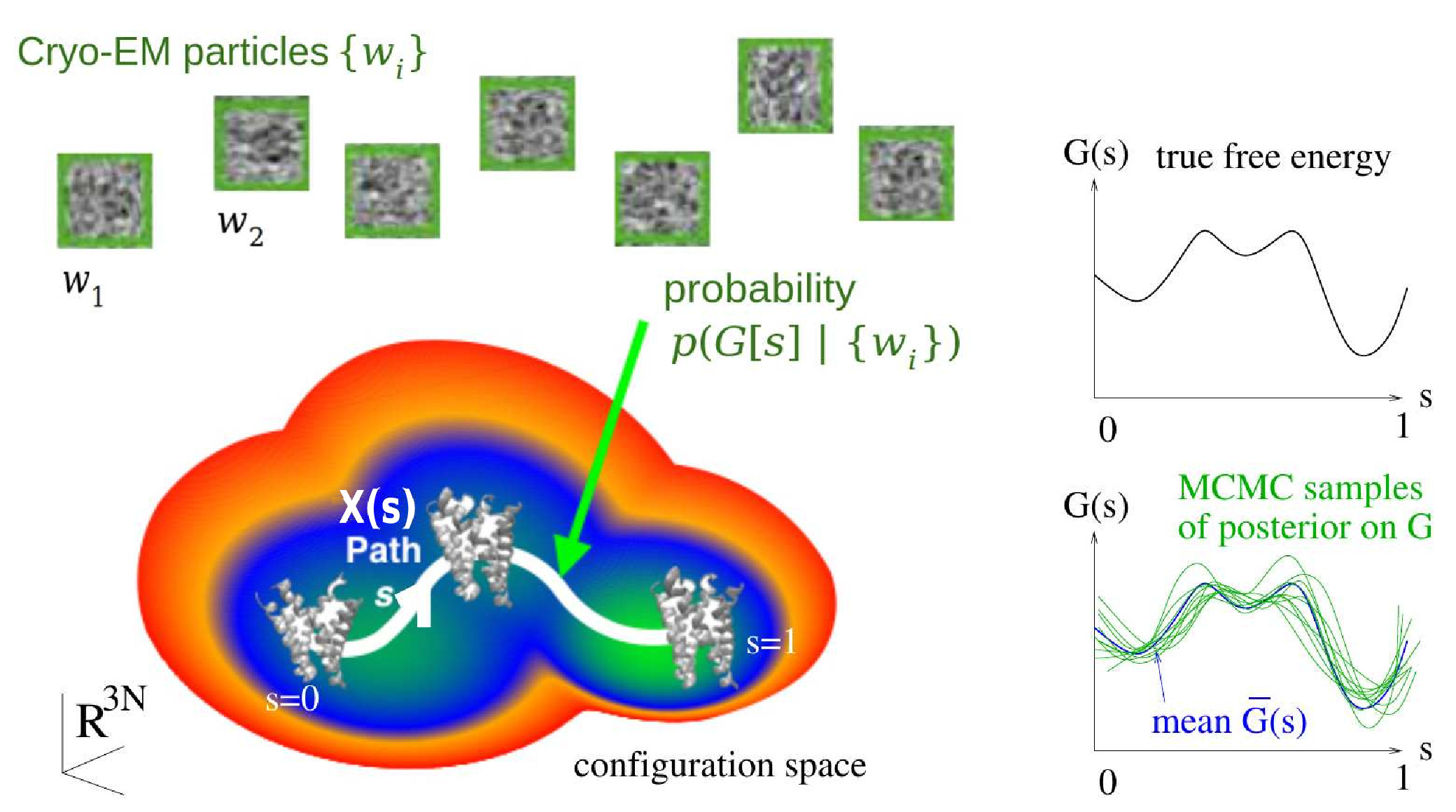}
\caption{{\bf Schematic representation of the path collective variable and Bayesian formalism for cryo-BIFE.} The main goal of our methodology is to determine the posterior probability distribution of free energy profiles $G(s)$ over a given configuration space path $X(s)$, given a set of noisy cryo-EM particle (projection) images $w = \{w_i\}$ from $i=1,...,I$. The green graphs on the right show independent samples drawn from this posterior, and the blue curve their mean. The black curve represents the true free energy. Variation arises from a detailed Bayesian model of imaging noise. The path $0\le s \le 1$ is discretized using $M$ nodes.}
\label{fig:path_sketch}
\end{figure}

\section*{Theory}

\subsection*{A path collective variable}
Consider a biomolecule of $N$ atoms. Inspired by ref.~\citep{Branduardi2007}, we will define a collective variable by projecting every possible molecular configuration onto a path in the biomolecule's configuration space.
We will use $x \in \mathbb{R}^{3N}$ to denote a particular configuration (conformation). We define the CV in a manner that allows for the extraction of a 1D free energy profile. 

Let a predetermined smooth 1D path $X$ in configuration space be parameterized by $0\le s\le 1$, so that $x=X(s)$ is a particular configuration (point on the path).
This path should span the relevant conformational changes of the system, and thermal motion should be relatively small in all directions transverse to the path. 
In Figure \ref{fig:path_sketch}, we show a schematic representation of the path $X$ (white curve) that connects the relevant metastable states (basins) in the conformational space.
At each configuration $x=X(s)$ one sets up transverse coordinates $z\in\mathbb{R}^{3N-1}$,
so that any configuration $x$ in a tubular neighborhood of the path may be written uniquely via a map $x={\cal X}(s,z)$, where $X(s)={\cal X}(s,0)$. This means that inverse functions $S(x)$ and $Z(x)$ exist such that ${\cal X}(S(x),Z(x)) = x$ for all $x$ in this neighborhood.
Our CV is defined by $S(x)$, \textit{i.e.} the parameter value $s$ of the unique point on the path nearest to a given thermally-accessible configuration $x$. For all points $X(s)$ on the path, $S(X(s))=s$ extracts their parameter.

In practice, one must discretize integrals (\textit{e.g.}, for the Bayesian analysis presented below) over $0\le s\le 1$. For this we use a simple $M$-node equispaced rule,
\begin{equation}
    \int_{0}^1 f(s) ds \approx \frac{1}{M} \sum_{m=1}^M f(s_m)~,
    \label{quad}
\end{equation}

applying to any smooth function $f$, where the parameter nodes are $s_m:=(m-1)/(M-1)$.
This defines a discrete set of 3D conformations (hereafter denoted as nodes) $x_m:=X(s_m)$, that take the system from a starting conformation $x_1$ to an ending one $x_M$.   Ideally, the parameterization of the path should have roughly uniform $|X'(s)|$, so that discrete conformations $x_m$ are approximately evenly spaced in $\mathbb{R}^{3N}$. However, constructing such a path might be challenging for many applications. In practice, one can assume that the cryo-EM images are likely to be generated by the nodes $x_m$. This assumption is commonly used for path-based algorithms for MD simulations \citep{Branduardi2007}, and for a well-chosen path it is formally justified by Laplace approximation in the low-temperature limit. 

The CV defined in reference \citep{Branduardi2007} compares 3D conformations (\textit{e.g.} from an MD trajectory) to the set of nodes belonging to the path $X$. Instead, in the following, we develop the cryo-BIFE method, a Bayesian formalism to calculate the posterior probability over free energy profiles $G(s)$, $0\le s\le 1$, along the predetermined path, given a set of cryo-EM images $w= \{w_i\}_{i=1}^I$.

\subsection*{The free energy profile along the path}

Here, we consider the biomolecule at thermal equilibrium. From Boltzmann statistics, the probability density at configuration $x \in \mathbb{R}^{3N}$ is given by
\begin{equation}
    \rho(x)=\frac{1}{Z}e^{- \beta H(x)} ~,
    \label{eq:boltzmann}
\end{equation}
where $H(x)$ is the system's Hamiltonian (potential energy of conformation $x$), and $Z=\int e^{- \beta H(x)} dx$ is the partition function. We now project this down to the CV. One may choose the map ${\cal X}(s,z)$ so that,
at each point on the path, $\frac{\partial x}{\partial z_j}$ for each of the transverse coordinates $z_j$, $j=1,\dots,3N-1$, are mutually orthonormal, and orthogonal to the path tangent vector $X'(s)$. Then, near to the path, the Jacobean of the map is the ``speed'' $|X'(s)|$ (note that $|z|^2$ then matches the squared-distance variable preferred in ref.~\citep{Branduardi2007}). 
A change of variables gives the marginalized probability density as
\begin{equation}
    \rho(s) = \int \delta\left(S(x)-s \right) \rho(x) dx =
    \frac{1}{Z} |X'(s)| \int e^{-\beta H({\cal X}(s,z))} dz
    \label{rhos}
\end{equation}
where $\delta$ is the 1D Dirac delta distribution, and in the last step we used equation \ref{eq:boltzmann} and the Jacobean.
Since only conformations near to the path are assumed relevant, the Jacobean here was taken as constant with respect to $z$. Note that the final integral in equation \ref{rhos} is a partition function restricted to the ``slice" transverse to $X$ at $s$.
It is then standard to interpret this $\rho(s)$ as the equilibrium density due to an
effective free energy profile (or Potential of Mean Force) $G(s)$ defined by
\begin{equation}
   \rho\left( s | G\right) = \frac{1}{Z_1}e^{-\beta G(s)}~,  \qquad  0\le s\le 1~,
    \label{eq:defn_fe_surface}
\end{equation}
a 1D analog of equation \ref{eq:boltzmann} with $Z_1 = \int e^{-\beta G(s)}ds$.
Our goal is to estimate the function $G$ from a large set of 2D cryo-EM images in a statistically rigorous fashion, up to an additive offset.

\subsection*{cryo-BIFE: a Bayesian approach for extracting the free energy profile using cryo-EM images}

In general, the underlying free energy for a system is unknown. However, in cryo-EM, we have access to a collection of (noisy) images $w := \{w_i\}_{i=1}^I$. Each image $w_i$ is a noisy unknown projection of the biomolecule with an unknown configuration $x$ that is independently distributed following equation \ref{eq:boltzmann}. In the CV approach we approximate this by a path configuration $x=X(s)$ where $s$ is a random variable distributed as in equation \ref{eq:defn_fe_surface}.

We propose a Bayesian approach to estimate the underlying $G(s)$ along the path. We will treat $G$ as a random variable, more precisely a random function on $[0,1]$, which will be conditioned on the experimental data. Applying Bayes' rule, we have
\begin{equation}
    p\left(G | w \right) = \frac{p\left(  w | G\right) p\left( G \right) }{p\left( w\right)}~,
    \label{eq:initial_Bayes}
\end{equation}
where $p\left(G | w \right)$ is the desired posterior density over free energy profiles induced by the observed data.
$p\left( w | G\right)$ is the sampling density of the set of observed images $w$ given a specific choice of free energy profile function $G$, also known as the likelihood. 
The term $p\left( G \right)$ is the prior density of the free energy profile.
The normalizing constant (also known as the evidence) $p\left(w\right) = \int p(G \mid w) \, p(w) \, \textrm{d}w$ may be ignored.
We assume that the cryo-EM images are conditionally independent given $G$, so that
\begin{equation}
    p\left(  w | G\right) = \prod_i p\left(  w_i | G\right),
\end{equation}
where $p\left(  w_i | G\right)$ is the sampling density of the single image $w_i$ given $G$.

Our imaging model involves two steps: drawing $s$ randomly according to $\rho$ in equation \ref{eq:defn_fe_surface}, then drawing a noisy image of the 3D molecular configuration $x=X(s)$ according to the full random set of imaging parameters (orientation, translation, etc). Because $s$ is an unknown (a.k.a. latent) variable, the likelihood of an image can be computed by {\em marginalizing} over $s$,
\begin{equation}
    p\left(w_i | G\right) \;=\; \int p(w_i | s) p(s|G) \, \textrm{d}s \;\approx\; \sum_m p(w_i | s_m)  p\left( s_m | G \right)~,
    \label{eq:like-1image}
\end{equation}
where the second step applies the quadrature approximation from equation~\ref{quad}, dropping the overall $1/M$ constant.
The second factor is simply the normalized equilibrium density, equation~\ref{eq:defn_fe_surface}, at the $m$th node, given the free energy profile $G$,
\begin{equation}
    p\left( s_m | G \right) =  \rho\left( s_m| G\right) ~.
    \label{eq:boltz-G}
\end{equation}
To evaluate the first factor in equation \ref{eq:like-1image}, the density in image space given the CV coordinate, we could consider the set of all molecular configurations in space that map to $s_m$. However, for many chemical systems and good reaction paths, this distribution is sharply peaked around the configuration on the path, $x_m$.
Thus, for our imaging model, we approximate
\begin{equation}
    p(w_i | s_m)\approx p(w_i | x_m)~.
\end{equation}

The cryo-EM imaging process is quite well understood, hence we can formulate an estimate of the sampling distribution $p\left( w_i | x_m  \right)$ (or likelihood as viewed as a function of $x_m$).
Considerable work has gone into constructing these likelihoods \citep{Scheres2012,Grigorieff2016,Scheres2008}. Here, we will use the BioEM formalism from ref. \citep{Cossio2013}, which involves a set of numerical marginalizations analogous to, but much larger in scale than, the one above, to calculate $p(w_i | x_m)$. See the Methods and refs. \citep{Cossio2013,cossio2017bioemgpu} for details about the BioEM calculations and their normalization.  We note that this method is not limited to the use of BioEM, and any likelihood formalism (\textit{e.g.}, those used for 3D reconstruction \citep{Scheres2012}) can be used.

Using the above expressions, and dropping normalization constants, the posterior over the entire set of images is given by
\begin{equation}
    p\left(G | w \right) \;\propto\; \prod_i  \sum_m p(w_i | x_m) \rho\left( s_m| G\right) p\left(G \right)
     ~.
    \label{eq:final_Bayes}
\end{equation}

In the Methods, we describe the Markov chain Monte Carlo (MCMC) method that we use to sample from equation (\ref{eq:final_Bayes}) and estimate the expected value of the free energy profile $G$ and its uncertainty given a predefined path and a set of particles. In the following, we validate and test cryo-BIFE over a diverse set of systems from a conformational change along one dimension, using synthetic images, to a membrane channel's calcium bound/unbound transition, using real cryo-EM data. 

\section*{Results}

To understand the effects of the physical parameters (\textit{e.g.}, involved in the image formation process) for recovering the free energy with cryo-BIFE, we designed several control systems where the projections are generated synthetically following the ideas of ref.\citep{PPR:PPR104388}. The first system consists of conformations of the Hsp90 chaperone representing a low-dimensional (1D-2D) conformational space. Then, the analysis is extended to a more realistic ensemble from converged MD simulations of the VGVAPG peptide, and an unconverged simulation of the SemiSWEET sugar membrane transporter~\citep{Latorraca2017}. Lastly, we apply cryo-BIFE over experimental cryo-EM data. For this propose, we have chosen raw images of TMEM16F a membrane channel and lipid scramblase~\citep{Feng2019} available at the EMPIAR databank~\citep{Iudin2016}. 

\subsection*{Free energy profile recovery over controlled datasets}

\subsubsection*{Hsp90 chaperone}

Hsp90 (a heat shock protein) is a chaperone involved in the folding process of several kinases, transcription factors and steroid hormone receptors \citep{Schopf2017}. This protein consists of two chains (A and B, containing 677 residues each) forming a V-like shape. Although Hsp90 is flexible, in the presence of certain ligands (\textbf{e.g.}, ATP) its conformational space can be reduced to a few degrees of freedom that go from an open to a closed state of the chains. Following the ideas described in ref.~\citep{PPR:PPR104388}, we reduce the open-closed dynamics of the Hsp90 into a one (1D) and two (2D) dimensional phase space where both chains are rotated in mutual normal directions and perpendicular to the symmetry axis (see the Methods). 

\subsubsection*{Free energy profile recovery for a 1D conformational change}

\begin{figure}[!ht]
\centering
\includegraphics[width=13cm]{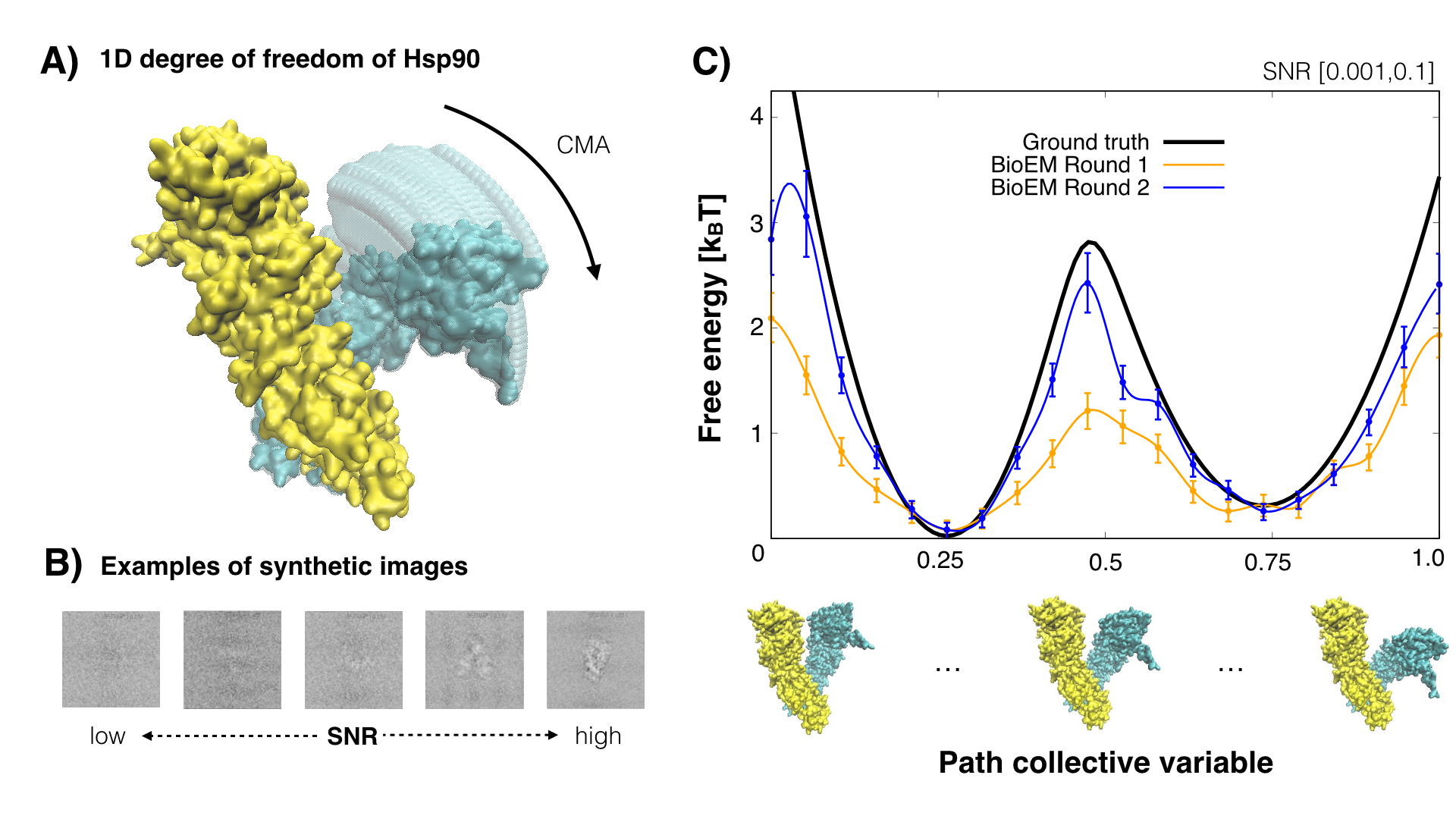}
\caption{{\bf 1D analysis of Hsp90.} A) Movement of Hsp90 along the single degree of freedom (CMA). The rotation of chain A relative to a fixed chain B. B) Examples of the synthetic images with varying SNR between [0.001,0.1]. C) Free energy profiles along the path for the entire set of images recovered from cryo-BIFE. The ground truth free energy is shown in black. The expected free energy using our formalism is shown for BioEM orientation rounds 1 and 2 in orange and blue, respectively. The R-hat test for the MCMC yielded 1.000 and 1.001 for BioEM round 1 and 2, respectively. The bars show the credible interval at 5\% and 95\% of the empirical quantile at each node. A cubic spline is used to fit the expected free energy profile, providing a smooth profile. The results show that a better estimate of the free energy is recovered for better orientation accuracy. }
\label{fig:Hsp90_1D}
\end{figure}

In Figure \ref{fig:Hsp90_1D}A, we show a 1D conformational change of Hsp90 determined by a rotation angle between its two chains, where chain B is fixed and chain A is rotated from the closed state to the open state (denoted by CMA). We define the path using twenty conformations equally spaced by $1^\circ$ in the rotation angle. The underlying synthetic free energy profile (\textit{i.e.} ground truth) along the path is shown as a black line in Figure \ref{fig:Hsp90_1D}C. We generated around 13300 synthetic images from the predetermined population of the twenty conformations (given by the Boltzmann factor of the ground truth free energy). The synthetic images have uniform random signal-to-noise-ratio (SNR) $\log_{10}([0.001,0.1])$, defocus [0.5,3] $\mu m$ and orientation angles (see the Methods). Examples of the synthetic particles are shown in Figure \ref{fig:Hsp90_1D}B.

To apply cryo-BIFE, we first precalculate the BioEM probabilities for the nodes along the path and all synthetic images for two BioEM rounds of orientation estimate (see the Methods). We apply the MCMC sampling strategy described in the Methods to extract the expected $\bar{G}(s)$ and the credible interval at 5\% and 95\% of the empirical quantile at each node. Figure \ref{fig:Hsp90_1D}C, shows the results of $\bar{G}(s)$ using all particles for the first and second BioEM rounds of orientation estimate. We note that the first round is less accurate than the second. This is also reflected in the recovery of the free energy profile $\bar{G}(s)$, where the second round has a much better performance. We conclude that the pose accuracy of the particles is crucial for extracting an adequate free energy estimate. The results from BioEM round 2 show that cryo-BIFE is able to recover the free energy profile for a wide range of SNRs and defocus. Interestingly, the credible intervals widen for higher free energy values, \textit{i.e.,} near the barrier, where there are less particles and the error is expected to be larger. 

\begin{figure}[!ht]
\centering
\includegraphics[width=\textwidth,height=\textheight,keepaspectratio]{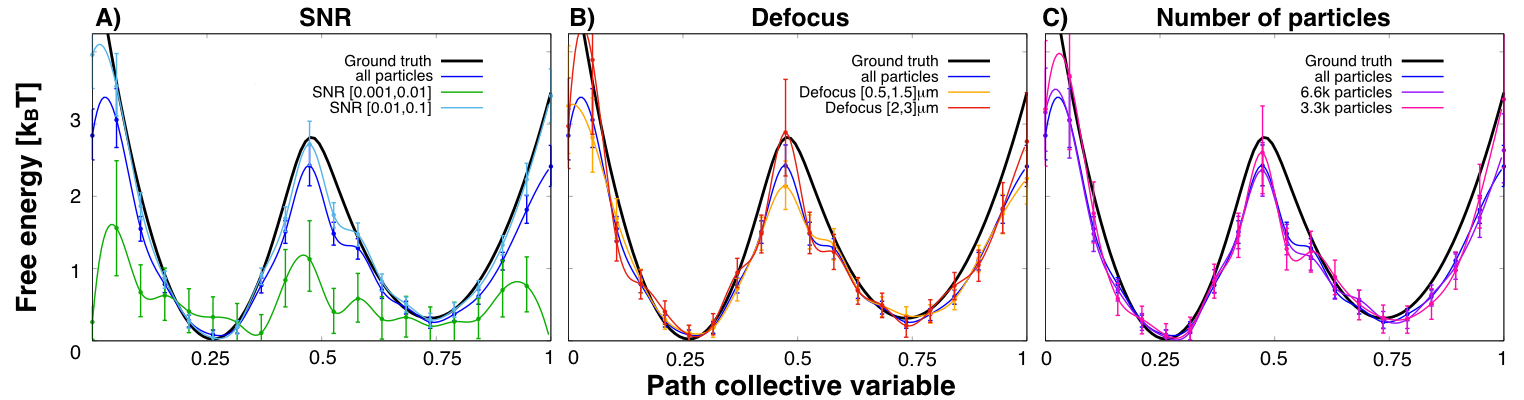}
\caption{{\bf  Cryo-BIFE free energy profile recovery for different cryo-EM conditions.} A) Particles grouped by SNR from [0.01,0.1] (cyan) and from [0.001,0.01] (green). Each subset contained around 6600 particles. B) Particles grouped by defocus. Sets with small defocus [0.5,1.5]$\mu m$ (orange) and large defocus [2,3] $\mu m$ (red). Each subset contained around 5300 particles. C) Particle subsets with different number of particles: 3300 (pink) and 6600 (purple). For reference, the ground truth and expected free energy profiles using all particles are shown in black and blue, respectively. The R-hat test for the MCMC yielded values $<1.01$ for all cases. The bars show the credible interval at 5\% and 95\% of the empirical quantile at each node. The results are for the second BioEM round of orientation estimate.  }
\label{fig:Hsp90_diff_conditions}
\end{figure}

We now study the performance of the method for different cryo-EM conditions. In Figure \ref{fig:Hsp90_diff_conditions}A, we divided the particle set in two: high SNRs from [0.01,0.1] and low SNRs from [0.001,0.01] with equal number of particles ($\sim$6600 each). The expected free energy calculated from cryo-BIFE is shown for the high and low SNRs sets (light blue and green, respectively) for the second BioEM orientation round. We also compare it to the $\bar{G}(s)$ using the entire set (blue line). We find a poor recovery for the low SNR set [0.001,0.01] and large errors, whereas the high SNR set behaves well. Interestingly, the free energy estimate for the entire particle set (SNR [0.001,0.1]) is slightly worse than for the high SNR set, but much better than for the low SNR set. The reason for this is that the Bayesian posterior (equation (\ref{eq:final_Bayes})) naturally weighs the contribution of each particle and particles with high SNR contribute much more weight to the posterior. 

In Figure \ref{fig:Hsp90_diff_conditions}B, we analyzed the effects of the defocus by grouping the particles with small defocus [0.5,1.5]$\mu m$ (orange line) and large defocus [2,3]$\mu m$ (red line). We find that the results for the large defocus are slightly better, but these have large errors around the barrier. We also studied how many particles were needed to recover the free energy profile. In Figure \ref{fig:Hsp90_diff_conditions}C, the results are shown for sets with 3300 (pink line) and 6600 (purple line) particles. Similarly to what has been found for 3D map validation \citep{Ortiz2020}, just a small set of particles ($\ge 3000$) randomly picked from the entire set is able to reproduce the underlying statistics. Contrary to 3D refinement, where large numbers of particles are required, our results indicate that conformational variability can be captured from a small set of particles. 

\subsubsection*{2D case Hsp90}

\begin{figure}[!ht]
\centering
\includegraphics[width=14cm]{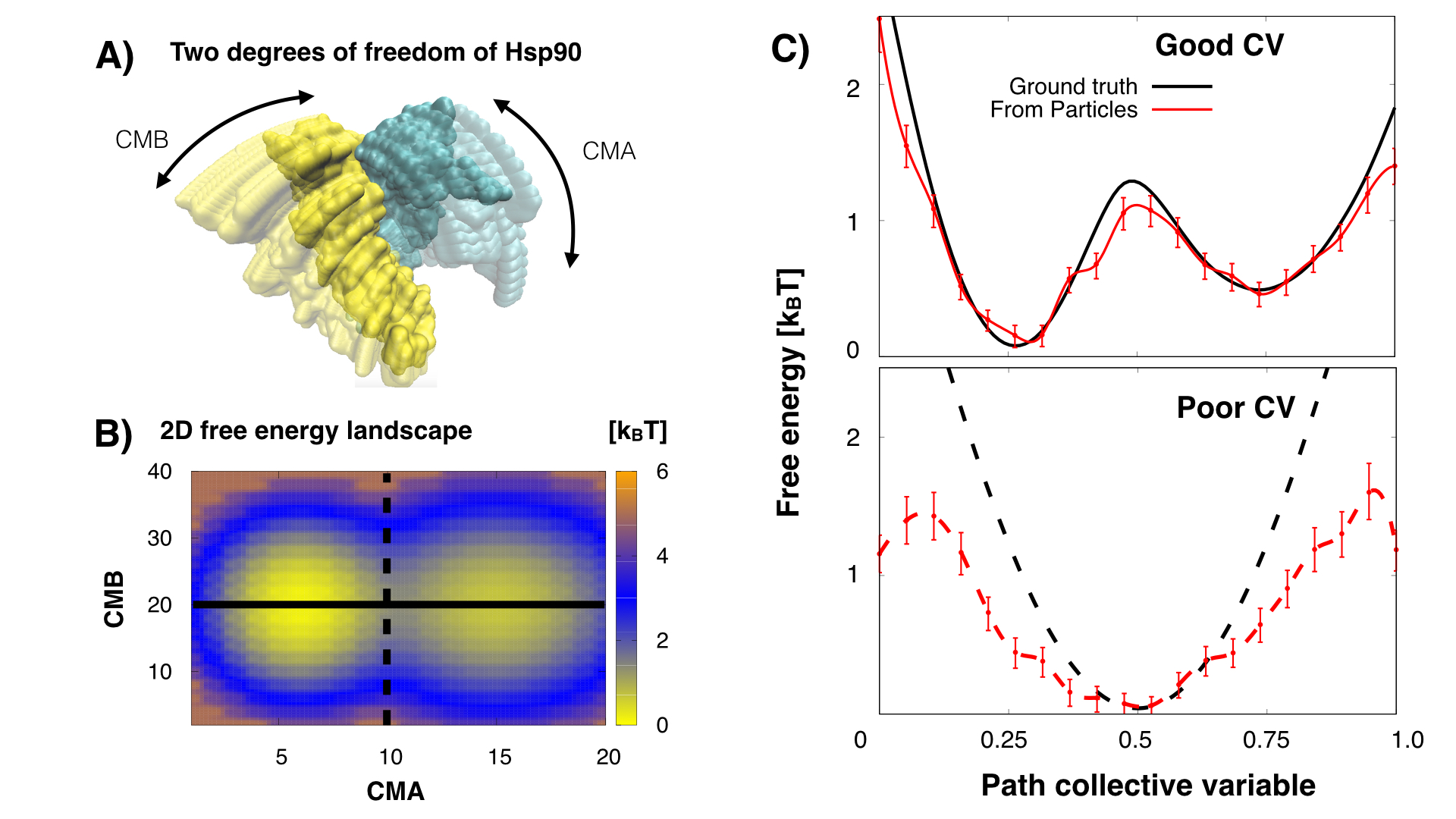}
\caption{\textbf{2D analysis of Hsp90.} A) Two degrees of freedom of Hsp90 along the CMA and CMB rotation directions (see the Methods). B) Ground truth free energy surface along CMA and CMB models. Black solid and dashed lines show a good and poor path CV, respectively. 
C) The free energy profiles along both path CVs, extracted with cryo-BIFE using synthetic particle images (red), are compared to the ground truth projected profiles (black). Solid and dashed lines are the results for the good and poor CVs, respectively. The R-hat test for the MCMC yielded values $<1.003$ for all cases. The bars show the credible interval at 5\% and 95\% of the empirical quantile at each node. The results are for the second BioEM round of orientation estimate.}
\label{fig:2D_Hsp90}
\end{figure}

As described in ref.~\citep{PPR:PPR104388}, Hsp90 is also characterized by second degree of freedom that is the rotation of chain B relative to the 1D rotation of chain A (see Figure \ref{fig:2D_Hsp90}A, and the Methods). We generated a synthetic 2D underlying free energy surface, shown in Figure \ref{fig:2D_Hsp90}B, with energy barrier of around 1$k_BT$ that is smaller than for the 1D case. Given the imagining conditions in cryo-EM experiments, we expect free energy barriers to be around this range. We generated 10000 synthetic particles, using the population given by the Boltzmann factor of ground truth free energy, with SNR [0.01,0.1], defocus [0.5,3]$\mu m$ and random orientations in $SO(3)$ (see the Methods).

To study the effects of the path CV, we defined two paths. The black solid line in Figure \ref{fig:2D_Hsp90}B shows a good path CV that passes along the relevant basins and the transition state of the system. In contrast, the black dashed line in Figure \ref{fig:2D_Hsp90}B shows a poor path CV. In Figure \ref{fig:2D_Hsp90}C, we compare the expected free energy profile extracted with cryo-BIFE to the ground truth (given by equation \ref{eq:defn_fe_surface}) along each path. We find good agreement between the underlying profile and the extracted free energy using the cryo-EM images along both paths. However, using the good path CV, we recover the meta-stable states of the system, the transition state and barrier height. These results indicate that if adequate paths are used, cryo-BIFE is valid for high-dimensional landscapes. On the contrary, if one uses a poor CV the barrier is lost and there is only one a single projected state. We note that this is an artifact of choosing a poor projection direction, and it is not a result of using 2D images. This highlights the importance of choosing an adequate path CV. 

\subsubsection*{VGVAPG system: a converged conformational ensemble}

\begin{figure}[!ht]
\centering
\includegraphics[width=14cm]{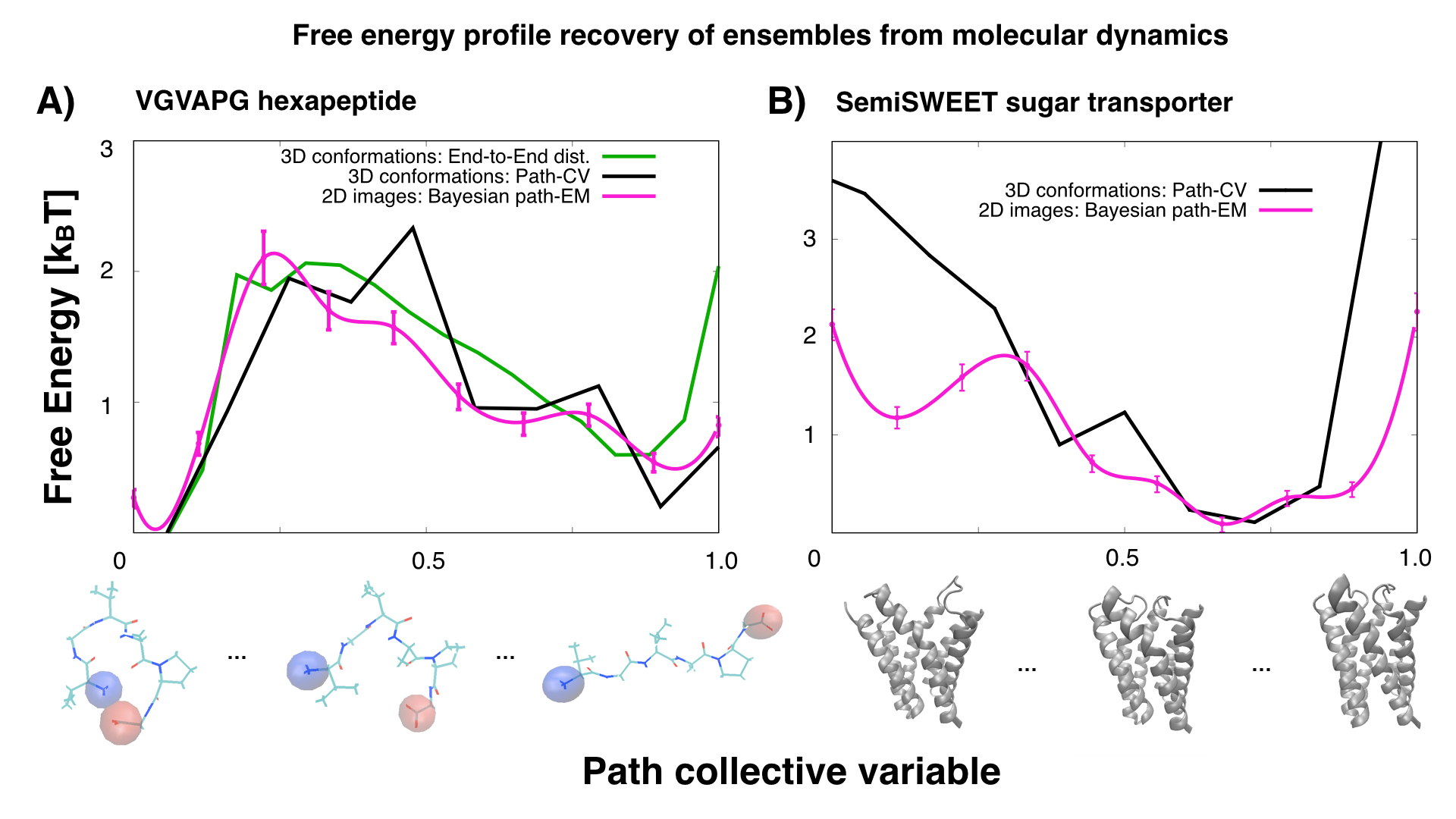}
\caption{{\bf Free energy profiles from 2D images (cryo-BIFE) or 3D conformations of the VGVAPG hexapeptide and semiSWEET transporter.} Free energy profile calculated over the 3D ensemble from MD using as CVs the end-to-end distance\citep{Donati2018} (green for the hexapeptide), and the path-CV with RMSD metric using equation 8 in ref. \citep{Branduardi2007} with $\lambda=50\AA^{-2}$ (black). We compare these profiles to the expected free energy $\bar{G}(s)$ extracted using cryo-BIFE with synthetic cryo-EM particles (pink line) for the VGVAPG hexapeptide (A) and the semiSWEET transporter (B). The R-hat test for the MCMC yielded values $<1.01$ for all cases. The bars show the credible interval at 5\% and 95\% of the empirical quantile at each node. See the Methods for details about the path and set of images for each system. }
\label{fig:VGVAPG}
\end{figure}

 MD simulations of the VGVAPG hexapeptide have been extensively used to test methods, such as Girsanov reweighting \citep{Donati2018}.  In the Supplementary Information, we present a video showing an example of the hexapeptide MD simulations performed for this work (see the Methods). The peptide has opposite charges in its extremes, which leads to conformational changes between an open state and closed state (see bottom Figure \ref{fig:VGVAPG}A). The underlying free energy is calculated using the 3D ensemble from the converged MD simulations. We used two CVs for the 3D conformations: the distance between the N-terminus and the C-terminus (which is considered good CV) and the 3D path-CV proposed by Branduardi \textit{et. al.} \citep{Branduardi2007} with the RMSD as metric (green and black lines, respectively, in Figure \ref{fig:VGVAPG}). Each CV is evaluated for each MD conformation, then a histogram is taken and the free energy is given via Boltzmann using the population of each histogram bin. 

We generated a set of 5688 synthetic images from the ensemble from the MD trajectory. The synthetic images had uniformly distributed random SNR, defocus and orientations (see the Methods). The path was created by selecting ten conformations with equally spaced end-to-end distance between successive nodes. We applied cryo-BIFE to extract the expected $\bar{G}(s)$ along the path (see the Methods). In Figure \ref{fig:VGVAPG}A, we compare the free energy profile from this analysis to the underlying free energy profiles calculated directly from the 3D ensemble from MD using the end-to-end distance CV or 3D path-CV \citep{Branduardi2007} for the same conformational path. The results show that with cryo-BIFE it is possible to recover the free energy profile using the 2D cryo-EM projections for a realistic ensemble. 

\subsubsection*{A membrane protein and small conformational changes}

The semiSWEET transporter is a membrane protein that transports sugar between cell membranes (Supplementary Figure 2A). Several unbiased MD simulations \citep{Latorraca2017} of this transporter were perform starting from the outward-open conformation. Eight trajectories showed a conformational change to the inward-open conformation. The C$_\alpha$- RMSD of snapshots of these trajectories to both states is shown in Supplementary Figure 2B. Although these simulations are not converged, we thought it interesting to investigate if cryo-BIFE is able to resolve the free energy profile of membrane proteins with nanodisk belts (as in the cryo-EM experiment), and small conformational changes ($<4$\AA~). We used these MD snapshots to generate a synthetic ensemble of semiSWEET conformations, which we use as reference. We generated 5280 synthetic images from this ensemble having each a nanodisk belt (see the Methods). To define the path, we clustered the MD conformations, and selected successive nodes that had quasi-equidistant difference in RMSD to the outward-opened state (see Supplementary Figure 2C). In Figure \ref{fig:VGVAPG}B, we compare the free energy profile from the 2D images to that from the 3D ensemble using Branduardi's 3D path-CV \citep{Branduardi2007}. We find relatively good agreement between the profiles around the minimum, however, for the high free energy regions the agreement is not as good. Nonetheless, these results encourage us to apply cryo-BIFE for studying the conformational transition between two states of a membrane channel using real cryo-EM particles.

\subsection*{Real cryo-EM data: TMEM16F ion channel} 

\begin{figure}[!ht]
\centering
\includegraphics[width=14cm]{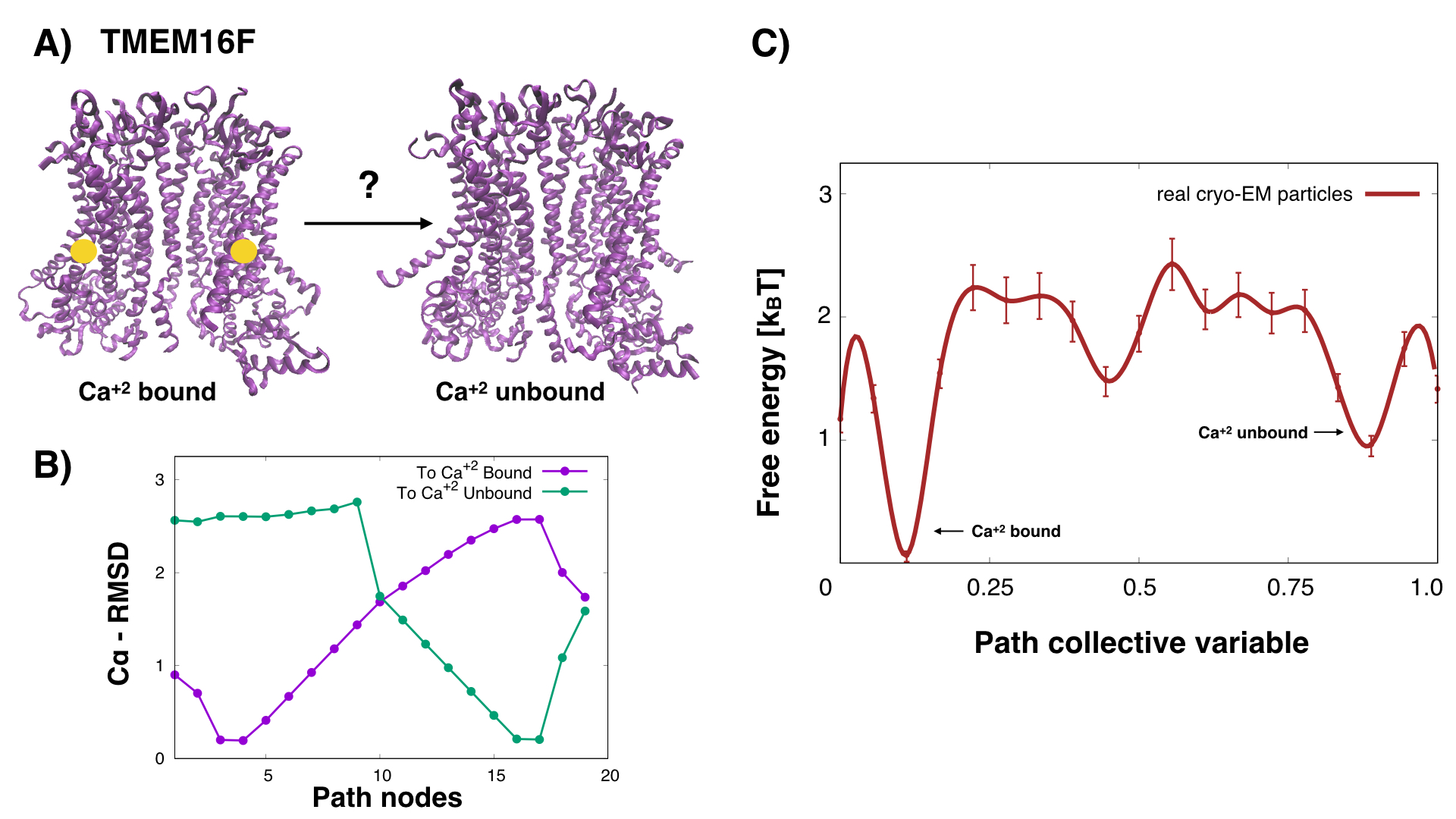}
\caption{{\bf Real cryo-EM data for studying the TMEM16F Ca$^{+2}$- bound/unbound transition with cryo-BIFE}. A) Ca$^{+2}$-bound to the Ca$^{+2}$-unbound states of TMEM16F (with PDB codes 6p46 and 6p47, respectively). B) C$_\alpha$ RMSD of the nodes along the path to the  Ca$^{+2}$-bound and Ca$^{+2}$-unbound states (purple and green, respectively). C) Free energy profile extracted along the path CV from real cryo-EM particles from the dataset used to generate the Ca$^{+2}$-bound reconstruction in digitonin\citep{Feng2019} (EMPIAR code 10278). The R-hat test for the MCMC yielded $1.001$. The bars show the credible interval at 5\% and 95\% of the empirical quantile at each node. Arrows point to the free energy basins corresponding to the Ca$^{+2}$-bound/unbound states.}
\label{fig:TMEM16F}
\end{figure}

TMEM16F is a membrane channel and lipid scramblase that is activated when calcium is bound. In ref. \citep{Feng2019}, cryo-EM experiments using different Ca$^{+2}$ conditions, and membrane/detergent compositions, were performed to resolve TMEM16F's Ca$^{+2}$ bound and unbound states. The cryo-EM particles at different conditions are available at the EMPIAR \citep{Iudin2016}. In this work, we focus on the dataset with around 1.2 million particles that was used to generate the Ca$^{+2}$-bound state in digitonin (EMPIAR code 10278). Since less than 15\% of these particles are used to generate the final reconstruction (all other particles are classified-out), we wanted to investigate \textit{i)} if there could be a small population of the  Ca$^{+2}$-unbound state in this set, and \textit{ii)} if we can extract a free energy profile from the Ca$^{+2}$-bound to the Ca$^{+2}$-unbound states. Starting from the PDB structures (Figure \ref{fig:TMEM16F}A), we used steered MD simulations, which included a lipid membrane and explicit solvent (see the Methods), to generate a path connecting both states. The C$_\alpha$-RMSD of the nodes to both states is shown in Figure \ref{fig:TMEM16F}B.  We randomly selected around 15000 particles from the entire set, \textit{i.e.}, not only those used for the final reconstruction. In Figure \ref{fig:TMEM16F}C, we show the expected free energy along the path using the same cyro-BIFE setup as for the previous systems. Interestingly, we find that both the Ca$^{+2}$-bound to the Ca$^{+2}$-unbound states correspond to metastable basins of the system. Because the cryo-EM data set was prepared with Ca$^{+2}$, it was expected that the Ca$^{+2}$-bound state corresponds to the lowest free energy minima. However, it is interesting that not all the particles belong to this state, and that the Ca$^{+2}$-unbound also has meta-stability. Remarkably, we find an additional basin around $s=0.45$, and the highest barrier is around $2.2k_BT$. These results show that it is possible to extract a reasonable and interesting free energy profile from real cryo-EM particles. 

\section*{Discussion}

In this work, we have developed cryo-BIFE, a methodology for extracting free energy profiles from cryo-EM experiments using a Bayesian approach with a path collective variable. The method was tested and validated over diverse systems ranging in complexity. Using controlled parameters, we found that the particle orientation accuracy and the SNR are important for adequately recovering the free energy profile. Our work is a proof of principal that, under reasonable cryo-EM conditions, it is possible to extract free energy profiles using individual cryo-EM particles.  

We have mostly focused on extracting the {\em expectation} of the free energy profile $G(s)$. However, our method produces (in the form of independent MCMC samples) the full posterior over such profiles, which contains much more information than just an average. In particular, it quantifies the degree of certainty with which $G(s)$ can be extracted given the noise in particle images. Credible intervals can be placed on any functional of $G$, such as downstream predictions (reaction rates, etc), simply by evaluating them for all $G$ in a set of MCMC samples. 

The method assumes that the choice of the configurational path is sufficiently good to describe the relevant metastable states and transitions of the system. However, this might not always be the case, as shown in Figure \ref{fig:2D_Hsp90}C. Although there are some methods to standardize the CV design~\citep{Sultan2018,Rogal2019}, determining if the path CV is adequate is challenging. Therefore, the development of path-optimization methods that determine the path that leads to the best free energy profile are still required. Moreover, for some systems, a single degree of freedom might be insufficient and extending the CV to multiple dimensions would be advantageous. 

It is important to note that the temperature plays a crucial role for extracting free energies. In principal, the flash-cooling process~\citep{Dubochet1988} is done sufficiently fast that the cryo-EM sample is trapped in the ensemble just before freezing. Consequently, the extracted free energy profile should be a representation of the system at that temperature. However, the freezing takes of the order of $\mu$s~\citep{Cabra2015}, so all relaxation processes faster than this time-scale are lost. It remains to be fully assessed how much does the freezing process affect the extracted free energy~\citep{Arsiccio2020}. On the other hand, to obtain high-resolution reconstructions, it is common to set the system at temperatures below the ambient one for over-stabilizing a single state. We hope that these methods to extract free energies will motivate the field to measure more at ambient temperature, and, moreover, use all particles (\textit{i.e.,} without having to discard large percentages).

All-in-all, extracting free energies from cryo-EM experiments, opens the field to the assessment of conformational dynamics from a biophysical perspective. By having the populations along relevant degrees of freedoms, the results go beyond the discussion of discrete versus continuous, and the biophysical mechanisms are truly revealed. Additional clues about the biomolecule's function are unraveled by the information of the metastable states (\textit{e.g.}, the size and shape of the free energy basins), of the activation barriers and of the location of the transition states of the system, as is common in single-molecule experiments.

\section*{Methods} 
\label{methods}

\subsection*{BioEM analysis}

The probabilities involved in equation (\ref{eq:final_Bayes}) are calculated using the BioEM algorithm \citep{Cossio2013}. Given an image $w_i$ and a 3D conformation (from a density map or atomic model) $x_m$, BioEM computes the probability $p(w_i | x_m)$ that $w_i$ is a projection of $x_m$. This probability is calculated by integrating the likelihood function $L(w_i\vert\Theta,x_m)$ (see the Supplementary Text) weighted by prior probabilities $p(\Theta)$  over all relevant physical parameters $\Theta$ for image formation (rotation angles, displacements, CTF parameters, noise variance,  normalization factor and offset \citep{Cossio2013,cossio2017bioemgpu}):
\begin{equation}
p(w_i | x_m) \propto \int L(w_i\vert\Theta,x_m)p(\Theta) d\Theta ~.
\label{eq:bioEM}
\end{equation}
The integrals over the noise variance, offset and normalization are done analytically, and all others are computed numerically, as described in ref.~\citep{cossio2017bioemgpu}. The prior probabilities of the orientation angles and the displacements are taken to be uniform over the integration interval. The prior for the CTF defocus parameter is a Gaussian distribution, its center and width will depend on the BioEM rounds described below. 

The BioEM orientational integral is divided in two stages referred to as Round 1 and Round 2, respectively. In BioEM round 1, $p(w_i | x_m)$ is calculated by integrating over a uniform orientation grid of 36864 quaternions, which is constructed following the method described in ref. \citep{Yershova2010}. The BioEM integration ranges and number of grid points for round 1 are presented in the Supplementary Text for each system. In BioEM round 2, a finer quaternion grid of 125 points is created around the ten best orientations (\textit{i.e.}, with highest probability) selected from BioEM round 1. In total, a 1250 quaternion grid is used for the second BioEM orientation round. For this round, the Gaussian prior for the defocus is centered at the synthetic/experimental value of each particle and its width is 0.3 $\mu m$. This procedure is similar at that described in refs.~\citep{Cossio2018,Ortiz2020}, however, here we calculate BioEM rounds 1 and 2 independently for each node of the path. We use the BioEM code from ref. \citep{cossio2017bioemgpu} with CPU and GPU acceleration. For one node along the path and 10000 particles of 128$\times$128 size, BioEM round 1 takes $\sim$ 6 hours on 24 CPU cores + 2 GPUs, and BioEM round 2 takes $\sim$ 3 hours on 24 CPU cores.

To apply the Bayesian formalism, given a predetermined set of nodes $x_m$ along the path, one just needs to calculate the BioEM probability $p(w_i | x_m)$ (equation (\ref{eq:bioEM})) once for each image and each node. Since the normalization constant is missing in equation \ref{eq:bioEM}, we normalize the BioEM probability such that $\sum_m p(w_i | x_m) =1$ for each $w_i$. Then, to estimate the expected free energy profile, one performs the MCMC algorithm described below.  

\subsection*{Markov chain Monte Carlo}

We use a Markov chain Monte Carlo (MCMC) method to draw independent samples of the free energy profile $G(s)$ from its Bayesian posterior defined in equation (\ref{eq:final_Bayes}). We have found that a standard Metropolis-Hastings algorithm, sampling the vector of values $\{G(s_m)\}_{m=1}^M$ at the discrete quadrature nodes, is adequate for our needs.
Initial values $G^0(s_m)$ are chosen independently and uniformly at random in $[-2,2]$ for $m=1,\dots,M$. Then, each MCMC step $i=1,2,\dots,N_{MC}$ comprises the following sub-steps:
\begin{itemize}
    \item We randomly select a node $m \in [1,M]$ with uniform probability. 
    \item We perform a random displacement of the free energy profile at the selected node $G^i(s_m) = G^{i-1}(s_m)+\delta g$ where $\delta g$ is uniformly randomly chosen in $[-0.5,0.5] k_{B}T$.
    \item We shift the free energy profile so that $\sum_m G^i(s_m)=0$.
    \item We evaluate equation (\ref{eq:final_Bayes}) using the samples $G^i(s_m)$ of this free energy, and the pre-calculated values of $\log(p(w_i | x_m))$ (described above in equation (\ref{eq:bioEM})) for all images and all nodes $m=1,\dots,M$. For the prior in equation (\ref{eq:final_Bayes}), we use $p(G)=\int \lambda e^{-\lambda\mathcal{G}} d\lambda=1/\mathcal{G}^2$, where $\mathcal{G}=\sum_{m=1}^{M-1}(G(s_{m+1})-G(s_{m}))^2$, which is a standard $\ell^2$ smoothness prior on the discrete differences, marginalized over the precision parameter $\lambda$.
    \item From this, the log-acceptance probability of the proposal is computed (here we omit $s$ for notation simplicity, so that $G$ may be thought of as a vector in $\mathbb{R}^M$):
    \begin{equation} 
    A (G^{i}, G^{i-1}) \; := \; \log\bigl(p(G^{i} | w )\bigr) - \log\bigl(p(G^{i-1} |  w )\bigr)~,
    \label{eq:acceptance}
    \end{equation}
    \item We choose a uniform random number $u \in [0,1]$. Then, if $\log(u) \leq  A (G^{i}, G^{i-1})$, the move is accepted, otherwise it is rejected (in which case $G^i =G^{i-1}$).
\end{itemize}
This procedure is iterated well beyond the time by which the distribution over samples has converged. For the systems analyzed in this work, we ran $R=8$ independent MCMC chains each with a total of $N_{MC}=200000$ steps. The expected value of the free energy at each node is calculated using all samples $i=1,\dots,R\, N_{MC}$, that is,
\begin{equation}
\bar{G}(s_m) = \frac{1}{R\, N_{MC}} \sum_{i} G^i(s_m)~.
\label{eq:Gmean}
\end{equation}
Finally, to recover an expected continuous function $\bar{G}(s)$, we fit a cubic spline through the values $\{\bar{G}(s_m)\}_{m=1}^M$ with knots being the nodes $s_m$. Since only free energy differences are relevant, we shifted $\bar{G}$ such that its minimum is zero.
The credible interval for each node is calculated at 5\% and 95\% of the empirical distribution. We perform the R-hat diagnostic test\citep{vehtari2021rank}, which compares the inter-chain variance to the variance within each chain, for monitoring lack of convergence in the MCMC using the arviz package \citep{arviz_2019}. R-hat values $\leq 1.1$ indicate convergence of the sampling.

The MCMC code is written in python3.5. It is optimized with the numba compiler, taking approximately 2 hours on 24 CPU cores for 13000 particles, 20 nodes, and 8 replicas each with 200000 MC steps. 

\subsection*{Synthetic particles}

We used a modification of the BioEM program \citep{cossio2017bioemgpu} to generate the synthetic cryo-EM particles following similar ideas as described in ref.~\citep{PPR:PPR104388} The image is created from a PDB structure (\textit{e.g.}, a conformation from the MD simulations) with a coarse-grained representation of the residues. Each residue is represented as a sphere with a corresponding radius and number of electrons \citep{Cossio2013}. The contrast transfer function (CTF) is modelled on top of the ideal image given a defocus, amplitude and B-factor (for details see the SI of ref. \citep{Cossio2013}). For the synthetic particles, the amplitude was 0.1 and the B-factor was $1\AA$. Gaussian noise is added on top of the CTF convoluted image. The standard deviation of the noise is determined (as in ref.~\citep{PPR:PPR104388}) using the SNR and variance of the image without noise (calculated within a circle of radius 40 pixels centered at the box center). All synthetic images were of box size 128$\times$128 pixels, however, the pixel size varied for each system.

\subsection*{Benchmark systems}

\subsubsection*{Hsp90 system}

The Hsp90 chaperone is a flexible protein involved in several biological processes related to protein folding \citep{Schopf2017}. When bound to certain ligands, its conformational landscape can be approximated by two relative motions of its chains (A and B)~\citep{PPR:PPR104388}. The Hsp90 dynamics is reduced to a 2D dimensional phase space, where both chains are rotated in mutual normal directions and perpendicular to the symmetry axis. In this work, we first assess conformations from just one degree of freedom (1D analysis), and then we assess images from conformations belonging to the 2D conformational space (2D analysis).

 To generate the conformations for the first degree of freedom (1D case), we started from the closed state (PDB ID 2cg9\citep{Ali2006}), removed the ATP ligand and residues 1-11 to avoid overlapping crashes. Chain B is fixed and chain A is rotated at $1^\circ$ steps around the center of mass of residues LEU674-ASN677, up to $20^\circ$ from the starting position, generating 20 conformations along this degree of freedom (denominated CMA motion~\citep{PPR:PPR104388}). These 20 conformations are used to the define the path for the 1D analysis (Figure \ref{fig:Hsp90_1D}A). Along this reaction coordinate, we propose a synthetic free energy (which determines the population occupancy) given by 
 $\exp(-\beta G_{true}(s))=\exp(-(19s-6)^2/8)+\exp(-(19s-15)^2/18)/3$ for $0\le s \le 1$. This ground truth free energy is shown as a black solid line in Figure \ref{fig:Hsp90_1D}C. Using this synthetic population for the conformations along the path, we generated 13333 synthetic images of pixel size $2.2$\AA~ with uniformly distributed random orientations in $SO(3)$, SNR in $\log_{10}[0.001,0.1]$ and defocus in [0.5,3] $\mu m$.
 
 For the 2D conformational landscape, we add a new rotation. Starting from each rotated chain A from the 1D case, residues ILE12-LEU442 of chain B are rotated in $2^\circ$ steps around the center of mass of residues LEU442-LEU443, in the normal direction to the plane generated by the 1D movement of chain A and the symmetry axis. This normal motion mode is referred to as CMB~\citep{PPR:PPR104388}. In total, 400 models are generated corresponding to $20\times 20$ rotations. We propose a 2D synthetic free energy given by 
  $\exp(-\beta G_{true}(u,v))=\exp(-(u-10)^2/18)(\exp(-(v-6)^2/8)/2+\exp(-(v-15)^2/24.5)/3.5)$ where $(u,v)$ are the CMA, CMB models respectively. These distributions are characterized by two minima localized at models $CMA=10,CMB=6,15$ separated by an energetic barrier of around $1k_BT$. We generated 10000 synthetic images of pixel size $2.2$\AA~ with uniformly distributed random orientations in $SO(3)$, SNR in $\log_{10}[0.01,0.1]$ and defocus in [0.5,3] $\mu m$. 
 For this case, we define two paths a good path CV with model $CMB=10$ fixed (equivalent to rotation angle $20^\circ$) and $CMA$ varying (solid line Figure \ref{fig:2D_Hsp90}B) and a poor path with $CMB$ varying and model $CMA=10$ fixed (dashed line Figure \ref{fig:2D_Hsp90}C). 

\subsubsection*{Converged 3D ensemble of the hexapeptide VGVAPG}

We use the conformational ensemble of the hexapeptide VGVAPG from a converged all-atom MD simulation in explicit solvent. We use GROMACS~\citep{Abraham2015} to perform a 230 ns MD simulation. The initial structure is extracted from the crystal structure of the Ca6 site mutant of Pro-SA-subtilisin~\citep{Uehara2012} with PBD code 3VHQ (residues 171 to 176)~\citep{Donati2018}. The peptide is solvated with a cubic water box, centered at the geometric center of the complex with at least 2.0 nm between any two periodic images. The AMBER99SB-ILDN~\citep{Lindorff-Larsen2010} force field and TIP3P water model are used~\citep{Jorgensen1983}. A minimization is done with the steepest descent algorithm and stopped when the maximum force is $\leq$ 1000 kJ/mol$\cdot$nm. Periodic boundary conditions are considered. We perform a 100 ps equilibration in an NVT ensemble using the velocity rescaling thermostat~\citep{bussi2007} followed by a 100 ps equilibration in an NPT ensemble using Parrinello-Rahman barostat\citep{parrinello1981}. The MD production run is performed without restraints, with a time step of 2 fs in an NPT ensemble at 300.15 K and 1 atm. We extracted MD snapshots (or frames) around every 40 ps, obtaining 5688 conformations (shown in Supplementary video 1).

We choose ten conformations to create the path, such that the nodes cover the relevant conformational changes of the system. This can be monitored using the end-to-end distance of the peptide, \textit{i.e.}, the distance between the nitrogen atom of the N-terminus, and the carboxyl carbon of the C-terminus~\citep{Donati2018}. We choose the nodes in such a way that the difference in the end-to-end distance between successive nodes is 1.8 $\AA$. We also use the end-to-end distance as a CV to extract the free energy profile. As a second CV, the 3D path-CV was calculated using RMSD between all the MD frames and the ten nodes belonging to the path with parameter $\lambda=50\AA^{-2}$ (using equation 8 of ref. \citep{Branduardi2007}). To calculate the free energy profile, we compute the value of each CV for all MD conformations, then we take the histogram (with number of bins equal to the number of nodes along the path) and the free energy is obtained using Boltzmann's factor with the population of the histogram-bins. 

From each MD conformation, we generate a synthetic image with pixel size of 0.3 \AA~  and with uniformly distributed random orientations in $SO(3)$, SNR in $\log_{10}[0.01,0.1]$  and defocus in [0.1,1.0] $\mu m$. Using the 5688 synthetic images and the same ten nodes of the path, we perform the cryo-BIFE analysis.

\subsubsection*{MD conformations of the SemiSWEET membrane transporter}

We use the conformations of the SemiSWEET membrane protein from eight unbiased MD trajectories that present a conformational change from the outward-open state to the inward-open state ~\citep{Latorraca2017}. The RMSD to both states is shown in Supplementary Figure 2B for 1761 snapshots of the trajectories taken every 1ns. Although the simulations have not converged, we take the conformations as a reference ensemble. We use the 3D path-CV  \citep{Branduardi2007} to calculate the reference free energy using the same parameters as described for the hexapeptide. To select the conformations of the path, we cluster the MD snapshots using the GROMACS \citep{Abraham2015} $g\_cluster$ tool with RMSD threshold of $1.3$\AA, generating 39 clusters. The representative conformation for each cluster (\textit{i.e.}, cluster center) os then compared to the outward-open and inward-open crystals using the RMSD. 12 cluster centers with quasi equidistant difference between successive frames of RMSD to the outward-open state were selected (see Supplementary Figure 2C). 

To generate the synthetic images, we use the SemiSWEET conformation together with a nanodisk (\textit{i.e.}, lipid) belt of $25\AA$ centered at the center of mass of the protein and extracted from the MD simulation. To coarse grain the nanodisk and imitate the effects of averaging, the heavy atoms of lipids were modelled with a 3\AA~radius and 10 electrons (see Supplementary Figure 2A). We generate three images from each MD snapshot (protein and nanodisk) of pixel size $0.7\AA$ with uniformly distributed random orientations in $SO(3)$, random SNR $\in \log_{10}[0.01,0.1]$ and random defocus $\in [0.5,1.5]\mu m$. A nanodisk belt was included also for each node for the BioEM comparison. The cryo-BIFE results are compared to the free energy extracted from the 3D MD conformations using the path CV \citep{Branduardi2007}. 

\subsection*{TMEM16F: experimental cryo-EM data}

\subsubsection*{Cryo-EM particles}

We use the cryo-EM particles of the TMEM16F membrane channel used to generate the calcium bound state \citep{Feng2019} from the EMPIAR dataset~\citep{Iudin2016} with code EMPIAR-10278. See ref.~\citep{Feng2019}, for the information about the experimental conditions. The images were recorded with a pixel size of $1.059~\AA$, box size of $256 \times 256$ pixels, with defocus values within the interval $[0.5,2.7]~\mu m$. For this work, we only use the $Ca^{+2}$-bound set (Digitonin\_Ca package), and we randomly select 20 micrographs from this set, resulting in around $15000$ images. Note that these images represent the entire set and not only those used for the final reconstruction. Since only 13\% of the particles, from the EMPIAR-10278 set, are used to create the $Ca^{+2}$-bound reconstruction\citep{Feng2019}, our hypothesis is that not all imaged particles belong to this state. Our aim is to extract a free energy profile from the $Ca^{+2}$-bound to the $Ca^{+2}$-unbound states using only the cryo-EM particles from the $Ca^{+2}$-added set. 

\subsubsection*{Steered MD for creating the TMEM16F path}

In order to generate the path, we use steered MD simulations from the $Ca^{+2}$-bound to the $Ca^{+2}$-unbound state. The simulations were performed as follows.  We started from the $Ca^{+2}$-bound structure (PDB ID 6p46). Since the structure has atoms missing, we added these using the Swiss model webserver~\citep{Waterhouse2018}. We note that because some residues have to accommodate to fit the missing residues the full atom structure is not identical to the PDB. Starting from the full atom model of 6p46, we added the membrane using CHARMM-GUI~\citep{Jo2008}, in a 3:1:1 ratio of 1-palmitoyl-2-oleoyl-sn-glycero-3-phosphocholine (POPC), 1-palmitoyl-2-oleoylsn-glycero-3-phosphoethanolamine (POPE), and 1-palmitoyl-2-oleoyl-sn-glycero-3-phospho-L-serine (POPS), respectively. A box size of $16.8076 \times  16.8076 \times 17.2012 nm$ was used with periodic boundary conditions and 122923 TIP3P water molecules were inserted. We used the GROMACS program~\citep{Abraham2015} with the CHARMM36M force field~\citep{Huang2017}. The temperature was controlled in the simulation by means of Berendsen termostat at 300 $K$, whereas the pressure was controlled with the Berendsen barostat at 1.0 atm~\citep{Berendsen1984}. The energy was then minimized using the steepest descent algorithm and stopped when the maximum force was $\leq$ 1000 kJ/mol$\cdot$nm. We used the leapfrog algorithm to propagate the equations of motion. The long-range electrostatic interactions are calculated using a PME scheme with a 1.2 nm cutoff. We performed two consecutive equilibrations, of 125 ps each, in an NVT ensemble with a time step of 1 fs. Then, we performed two equillibrations in an NPT ensemble, where the first was of 125 ps and time step of 1 fs, and the last was of 1.5ns, with a time step of 2 fs. For the equilibration in the NPT ensemble, the pressure coupling was of semi-isotropic type. The backbone atoms of the protein were restrained throughout the equilibrations.

After the MD equillibration, we performed steered MD simulations~\citep{Grubmller1996} using the GROMACS program~\citep{Abraham2015} patched with the PLUMED 2.5 library~\citep{Tribello2014}. A first target structure for the steered MD is the $Ca^{+2}$-unbound state (PDB ID 6p47). We use the RMSD of the C$_\alpha$ atoms to steer the dynamics between the initial structure and the target structure. The steering harmonic potential has an initial force constant of $5000$ and ending at $260000\, kJ/mol/nm^2$. We noticed that a threshold of 0.2~\AA~in RMSD to the $Ca^{+2}$-unbound reference is reached very quickly, in less than $1ns$ (Supplementary Figure 2). A second steered MD simulation was needed in order to go from the initial system (all atom system) to the 6p46 PDB structure. This steered MD used the same parameters mentioned before. We also ran two short ($1ns$) unbiased MD simulations starting from each state (\textit{i.e.}, closest conformation to PDB 6p47 and 6p46). These trajectories allowed us to build a path from the $Ca^{+2}$-bound to the $Ca^{+2}$-unbound states. We used the C$_\alpha$-RMSD to the $Ca^{+2}$-bound state to select 19 nodes, where successive nodes are as equidistant as possible (see Figure \ref{fig:TMEM16F}B). To mimic the detergent in the cryo-EM images, we included a membrane nanodisk, for these nodes, of 50\AA~ radius taken from lipids in the MD simulations and centered at the center of mass of the protein. The nanodisk was modelled in a coarse-grained manner, similarly as for the SemiSWEET transporter.

\section*{Data availability}
The BioEM code is available at
https://github.com/bio-phys/BioEM.

\section*{Acknowledgements}
J.G-B., S.O. and P.C. were supported by MinCiencias, Ruta N, University of Antioquia, Colombia, and the Max Planck Society, Germany. The Flatiron Institute is a division of the Simons Foundation. The authors also acknowledge: Naomi Latorraca, Ron Dror for the availability of the MD trajectories; Cristian Rocha for help setting up the TMEMF16F membrane; and Johans Restrepo, Yifan Cheng, Ahmad Reza Mehdipour and Gerhard Hummer for useful discussions.

\section*{Author contributions statement}

J.G-B., S.O. and P.C. developed the concept, and performed the BioEM analysis. K. P-R performed the MD simulations of the hexapeptide. E.H.T., B.C., A.H.B. and P.C. developed the theory and methods. All authors contributed to all figures, wrote and reviewed the manuscript.

\section*{Additional information}

The authors declare no competing interests. Materials, data and associated protocols are promptly available for all readers.



%% References
%% Usage of \citep is as follows:
%%   \citep{key}          ==>>  [#]
%%   \citep[chap. 2]{key} ==>>  [#, chap. 2]
%%   \citept{key}         ==>>  Author [#]



%% New version of the num-names style
%\bibliographystyle{model2-names}
%\bibliographystyle{elsarticle-harv}\biboptions{comma,authoryear}
%\bibliography{ref}




%%
%% End of file `elsarticle-template-1-num.tex'.

\end{document}